\documentclass[a4paper,11pt]{article}
\pdfoutput=1 

\usepackage{jinstpub} 

\usepackage{lineno}

\usepackage{multirow}
\usepackage{wasysym}
\usepackage{xcolor}
\usepackage{xspace}
\usepackage{subcaption}
\usepackage{graphicx}
\usepackage[utf8]{inputenc}
\usepackage{makecell}
\usepackage{upgreek}
\usepackage[export]{adjustbox}
\usepackage{siunitx} 
\DeclareSIUnit\sq{\ensuremath{\Box}}                           
\usepackage[rawfloats=true]{floatrow}
\usepackage{xcolor}

\usepackage{hyperref}
\usepackage{orcidlink}
\newcommand{\FCCee}{\ensuremath{\textrm{FCC}_{ee}}\xspace}

\title{Simulations of MAPS designs for FCC-ee}

\author[a]{Danush Shekar\note{Corresponding author.}}
\author[b]{Doug Berry}
\author[b]{Ron Lipton}
\author[a]{Corrinne Mills}

\affiliation[a]{University of Illinois at Chicago, Chicago, IL 60607, USA}
\affiliation[b]{Fermi National Accelerator Laboratory, Batavia, IL 60510, USA}

\emailAdd{dsheka3@uic.edu}

\abstract{Future electron-positron colliders such as the future circular collider require vertex detectors with excellent spatial resolution and a low material budget.
Monolithic active pixel sensors (MAPS) are a leading candidate technology because they integrate the sensing volume and readout electronics within a single thin silicon substrate and can operate at power densities compatible with air cooling.
This paper presents the design and optimization of a high-voltage MAPS (HV-MAPS) pixel structure with a thickness of 25~\unit{\micro\meter} and a pitch of 50~\unit{\micro\meter}, developed using TCAD process and device simulations.
The implantation energies and doses of the top-side implants are optimized with respect to the maximum electric field, sensor capacitance, and signals induced at each contact by minimum-ionizing particles.
The optimized design reaches full depletion at a bias voltage of approximately $-3$~\unit{V} and has an estimated breakdown voltage magnitude of 300--390~\unit{V}.
At a bias voltage of $-100$~\unit{V}, the simulated charge-collection efficiency is approximately 90\% outside the well regions and 80\% within them and remains nearly independent of the angle of incidence.}


\begin{document}
\maketitle
\flushbottom

\section{Introduction}\label{sec:introduction}

The high-energy physics community is actively pursuing plans to construct a future electron-positron ($e^{+}e^{-}$) collider for precision studies of the Higgs boson.
One of the most developed proposals is the Future Circular Collider (\FCCee) at the European Organization for Nuclear Research (CERN).
If approved, construction of the \FCCee could begin in the 2030s, with operation foreseen to begin in the mid-2040s.
The facility will have a circumference of approximately 91~\unit{\kilo\meter}~\cite{FCC:2025lpp}.
Its physics program will focus on precision measurements of the Higgs boson, the W and Z bosons, and the top quark using collisions at center-of-mass energies ranging from 88 to 365~\unit{\giga\eV}.

This physics program places stringent requirements on the detector systems, particularly the tracking and vertexing subsystems.
Precise momentum measurements, secondary-vertex reconstruction, and flavor tagging are central to measurements of Higgs and electroweak observables, as well as to studies of heavy-flavor and tau final states.
Consequently, tracking in this collider environment must provide excellent single-hit spatial resolution while minimizing detector material and systematic uncertainties to preserve performance in the presence of beam-induced backgrounds and high event rates.
To meet these requirements, several proposed detector concepts~\cite{Ilg:2025zri, Bacchetta:2019fmz, allegro-detector} incorporate silicon-based vertex detectors close to the interaction point.
Recent developments in low-mass, low-power CMOS monolithic active pixel sensors, together with their potential for large-area coverage, make this technology a promising candidate for vertex and tracking detectors at \FCCee.

MAPS technology is foreseen for use in the vertex detectors of \FCCee detector concepts because it integrates the sensing element and readout electronics within a single silicon substrate, enabling very thin detector layers and low-power operation.
The low power consumption makes air cooling feasible, thereby contributing to a low material budget.

The remainder of this paper is organized as follows.
Section~\ref{sec:maps-overview-design} reviews the requirements for MAPS at \FCCee and describes the design and optimization of the HV-MAPS structure developed to address these requirements.
The implant energies and doses are optimized with respect to the maximum electric field, sensor capacitance, and signals induced at each contact by minimum-ionizing particles (MIPs).
Section~\ref{sec:hvmaps-performance} presents the simulated performance of the optimized HV-MAPS design.
Finally, Section~\ref{sec:conclusion} summarizes the results and provides an outlook on future work.

\section{MAPS requirements and HV-MAPS design}
\label{sec:maps-overview-design}

Several MAPS technologies have demonstrated performance relevant to the \FCCee environment.
Existing developments, including ARCADIA~\cite{Neubuser:2022wdv, Rignanese:2026ooo}, MOSAIX~\cite{SIDDHANTA2026100574, Menzel:2948541}, and EIC-LAS for ePIC~\cite{Turrisi:2026fgt,Glover:2025set}, demonstrate detector designs capable of meeting demanding particle-tracking requirements.
While these technologies are primarily being developed for collider environments with higher particle densities and radiation levels than expected at \FCCee, further optimization is possible to exploit the comparatively benign radiation environment and stringent material-budget requirements of \FCCee.
A MAPS detector for \FCCee should provide a material budget below $0.2\%~X_{0}$ per layer, a spatial resolution better than 3~\unit{\micro\meter}, and an angular resolution of approximately 0.1~\unit{\milli\radian}~\cite{Barchetta:2021ibt}.

The current target resolution for the transverse impact parameter is approximately
\begin{equation}
3~\unit{\micro\meter}\bigoplus\frac{15~\unit{\micro\meter\giga\eV}}{p\sin^{3/2}\theta},
\end{equation}
as described in Refs.~\cite{FCC:2025lpp, Ilg:2025iaj}.
A low material budget and high spatial resolution are also essential for maintaining the required primary- and secondary-vertex reconstruction performance while minimizing degradation from multiple scattering.
Ongoing \FCCee R\&D efforts are focused on adapting MAPS technology to the specific collider environment, with emphasis on high spatial precision, improved rate capability, increased radiation tolerance, and, in some cases, fast timing performance.

Several technical approaches are being pursued to adapt MAPS designs to these requirements.
These include reducing charge-collection times through larger or more favorable collection electrodes, optimizing epitaxial-layer properties and transistor choices for improved radiation tolerance, reducing inactive edge regions, and increasing front-end speed while maintaining power densities compatible with air cooling and lightweight mechanical support structures.

Current state-of-the-art MAPS detectors typically involve trade-offs between material budget, power consumption, and operation at the hit rates of up to 100~\unit{\mega\hertz}/\unit{\centi\meter\squared} expected for the \FCCee Tera-Z program.
An ideal vertex detector would therefore be very thin, support high-rate operation with a low thermal load, and remain radiation tolerant up to approximately 100~\unit{\kilo\gray}.


\subsection{HV-MAPS design}

\begin{figure}[hbtp]
    \centering
    \includegraphics[width=0.49\textwidth]{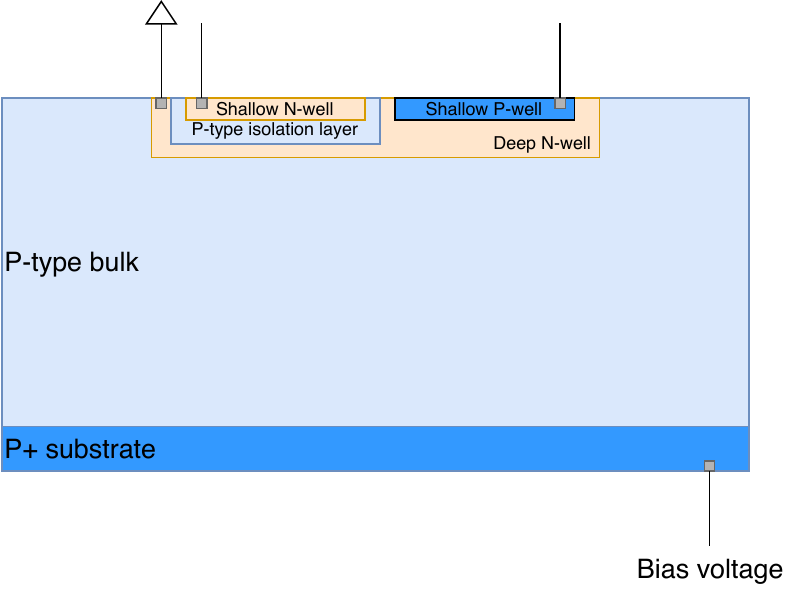}
    \includegraphics[width=0.49\textwidth]{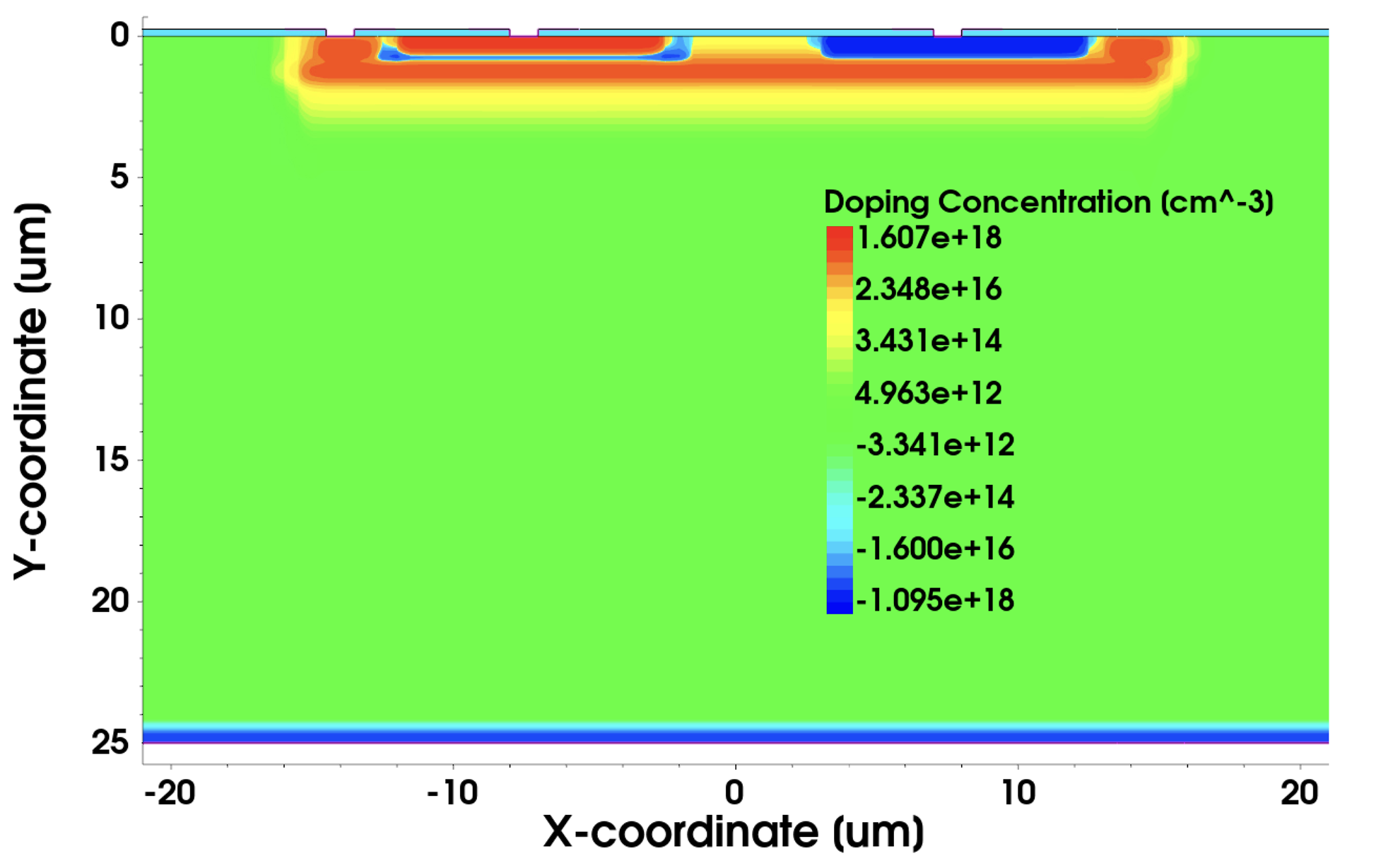}
    \caption{(left) Schematic of the sensor design and (right) doping concentration profile of a single pixel.}
    \label{fig:schematic-and-doping-profile}
\end{figure}

A MAPS sensor with a thickness of 25~\unit{\micro\meter}, a pixel pitch of 50~\unit{\micro\meter}, and a bulk resistivity of 13.3~\unit{\kilo\ohm\cdot\cm} was studied.
A schematic of a single-pixel unit cell, together with the corresponding doping concentration profile, is shown in Figure~\ref{fig:schematic-and-doping-profile}.
Process and device simulations were performed using Synopsys Sentaurus Technology Computer-Aided Design (TCAD)~\cite{Wu2018} for a two-dimensional structure consisting of three pixels.
The structure features a p++ backplane, through which the bias voltage is applied, and four primary implants on the top side: a shallow p-type well (SPwell), a shallow n-type well (SNwell), a deep n-type well (DNwell), and a p-type isolation layer (Piso) separating the two n-type wells.
At present, the full CMOS circuitry is not included in the TCAD device simulations, and simplified implant structures are used in its place.
The full circuitry will be incorporated once the design is finalized and before submission to the foundry.

The current design targets operation at high bias voltage to accommodate the expected radiation environment at \FCCee.
This is enabled by the Piso and DNwell layers, which isolate the shallow implants from high electric fields and currents induced by charged particles traversing the sensor.
Simulations were performed to optimize the top-side design parameters, specifically the implantation energies and doses.
The implantation energy and dose of each implant were varied independently in a series of parameter scans to assess their individual effects on device performance.
Several performance metrics were used in this assessment:
\begin{enumerate}
\item The maximum electric field magnitude within the silicon volume, to ensure operation below the sensor breakdown threshold.
\item The electric field magnitudes within the shallow implants, to avoid excessive electric fields in the sensitive circuitry that these implants will contain.
\item The total sensor capacitance and inter-implant capacitance, which should be minimized to reduce noise when coupled to the readout chain.
\item The signals induced at all contacts by MIPs, with the goal of maximizing charge collection at the DNwell contact while minimizing charge collection at the shallow-implant contacts.
\end{enumerate}
For the transient simulations described in the final item, a MIP is propagated through the center of a pixel and deposits approximately 80 electron-hole pairs per \unit{\micro\meter}.

\begin{figure}[hbtp]
    \centering
    \includegraphics[width=0.32\textwidth]{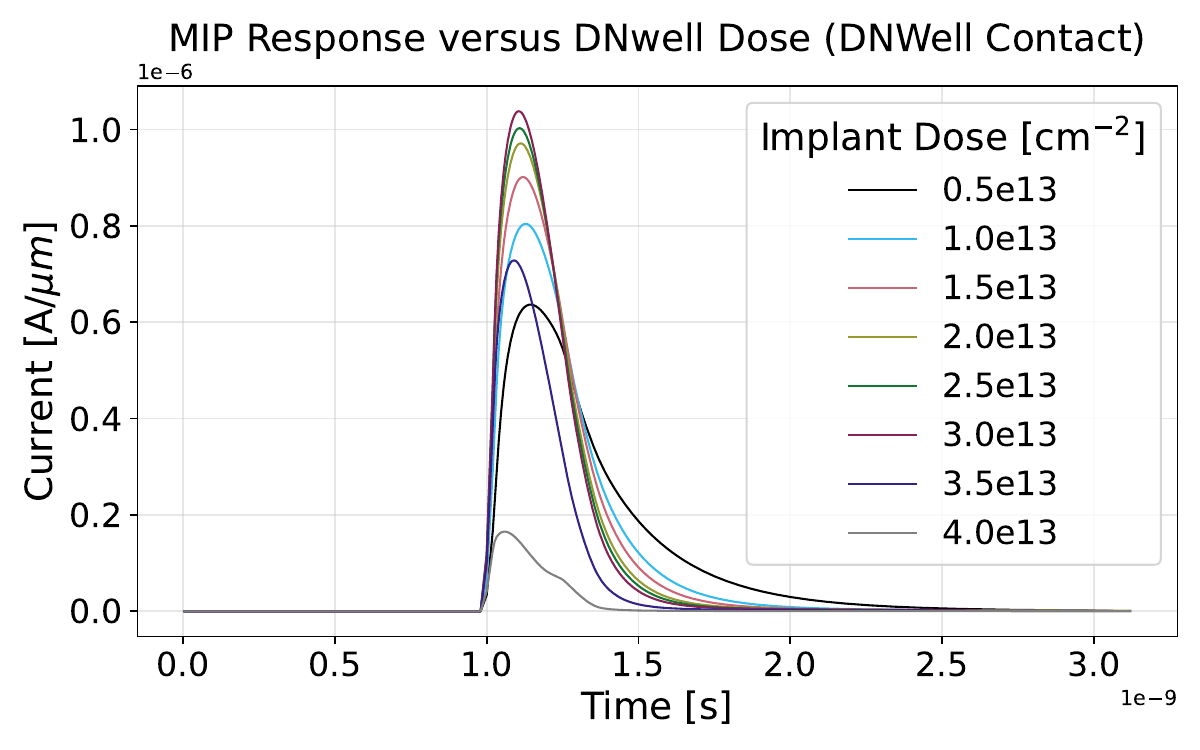}
    \includegraphics[width=0.32\textwidth]{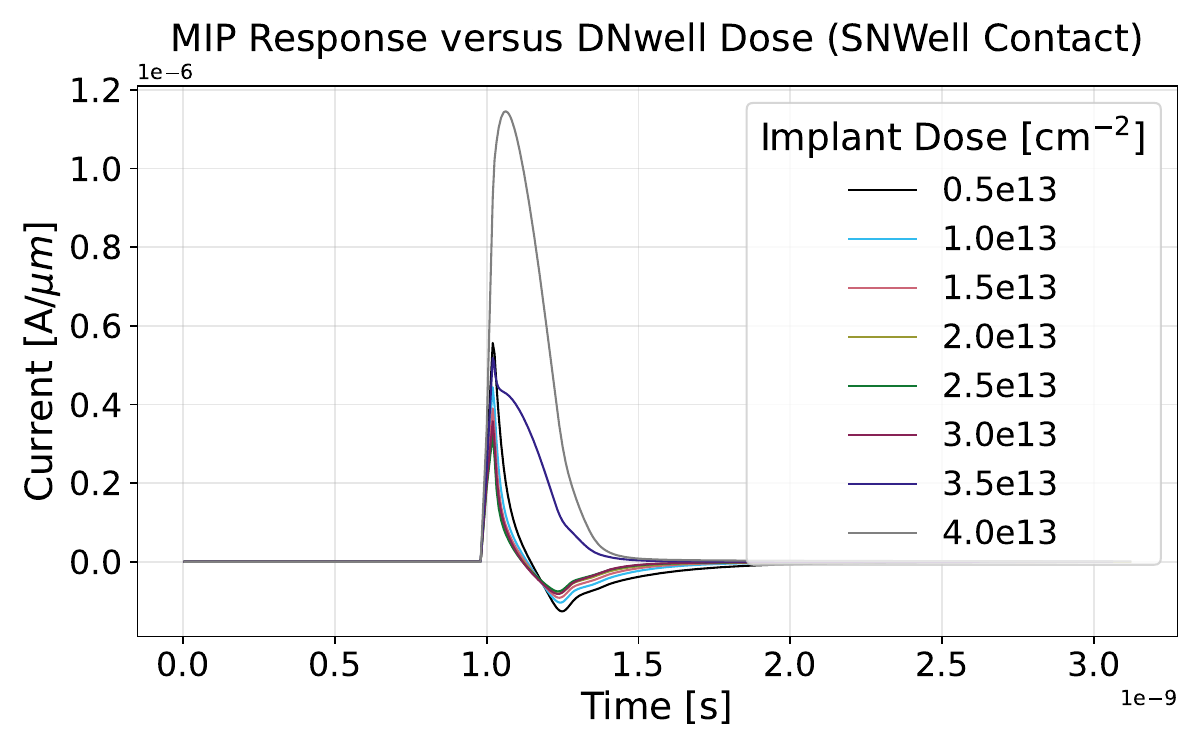}
    \includegraphics[width=0.32\textwidth]{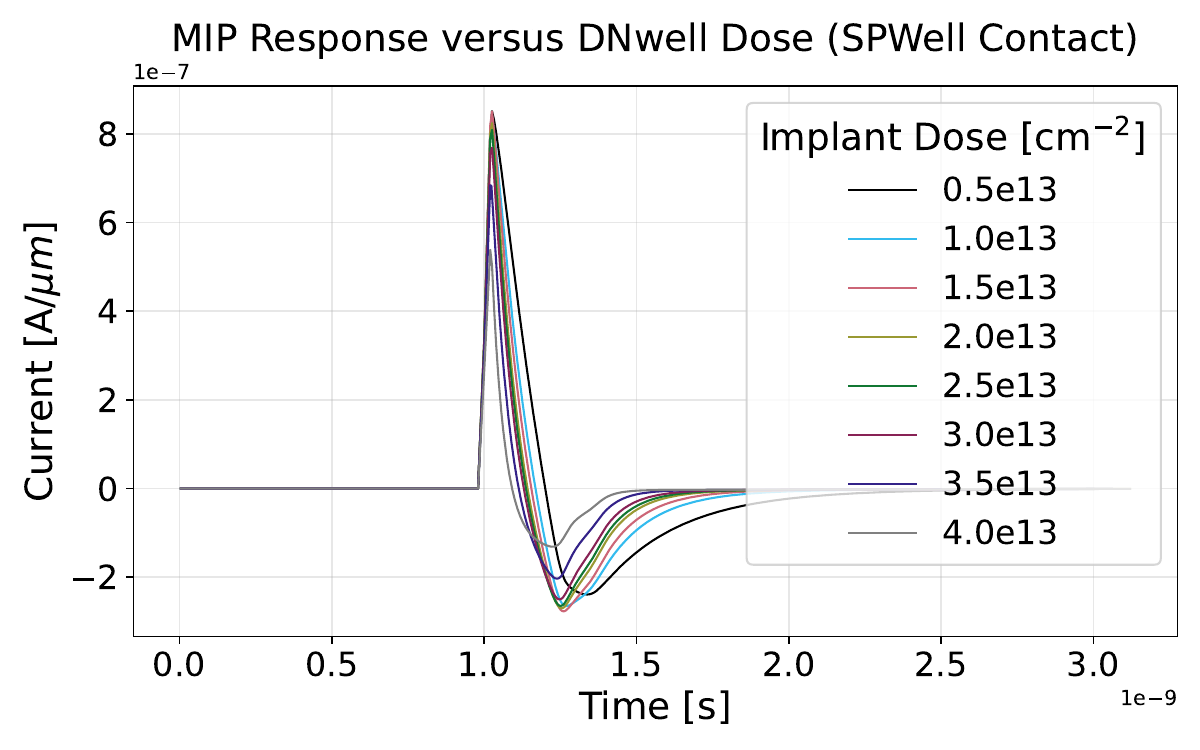}
    \caption{Response of the (left) DNwell, (center) SNwell, and (right) SPwell contacts to normally incident MIPs traversing the center of the pixel, shown as a function of the DNwell implantation dose.}
    \label{fig:mip-response-vs-dose}
\end{figure}

\begin{figure}[hbtp]
    \centering
    \includegraphics[width=0.32\textwidth]{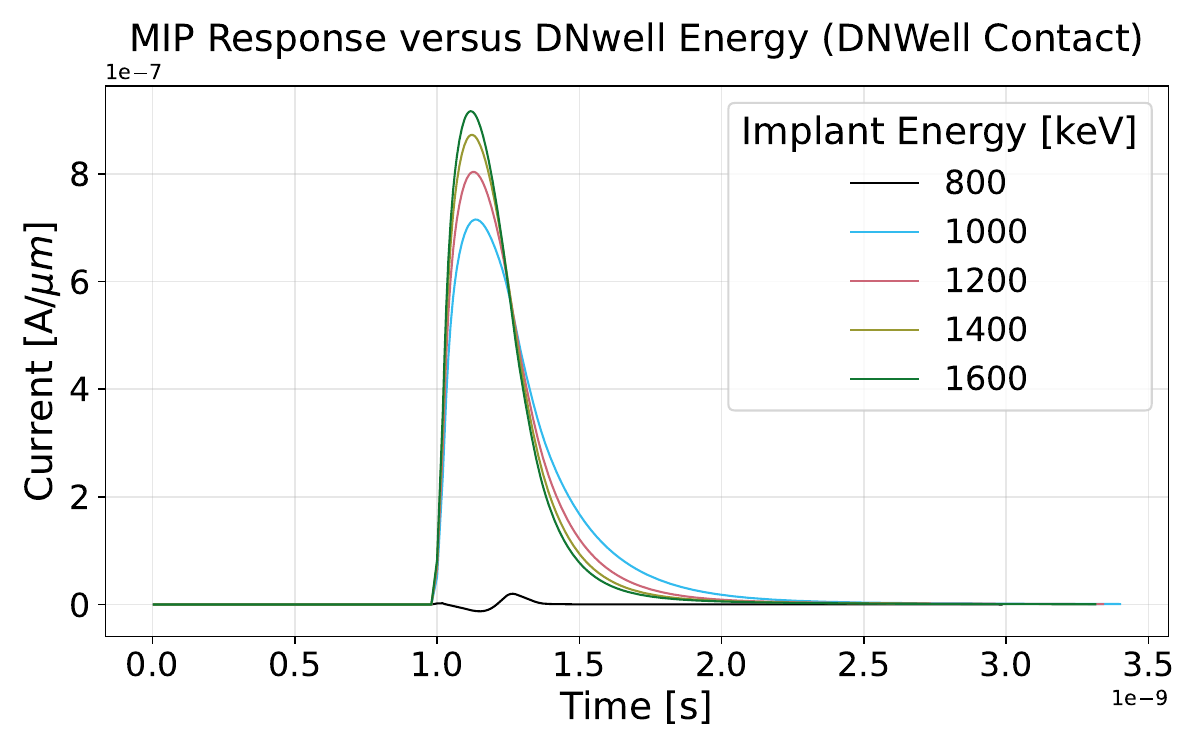}
    \includegraphics[width=0.32\textwidth]{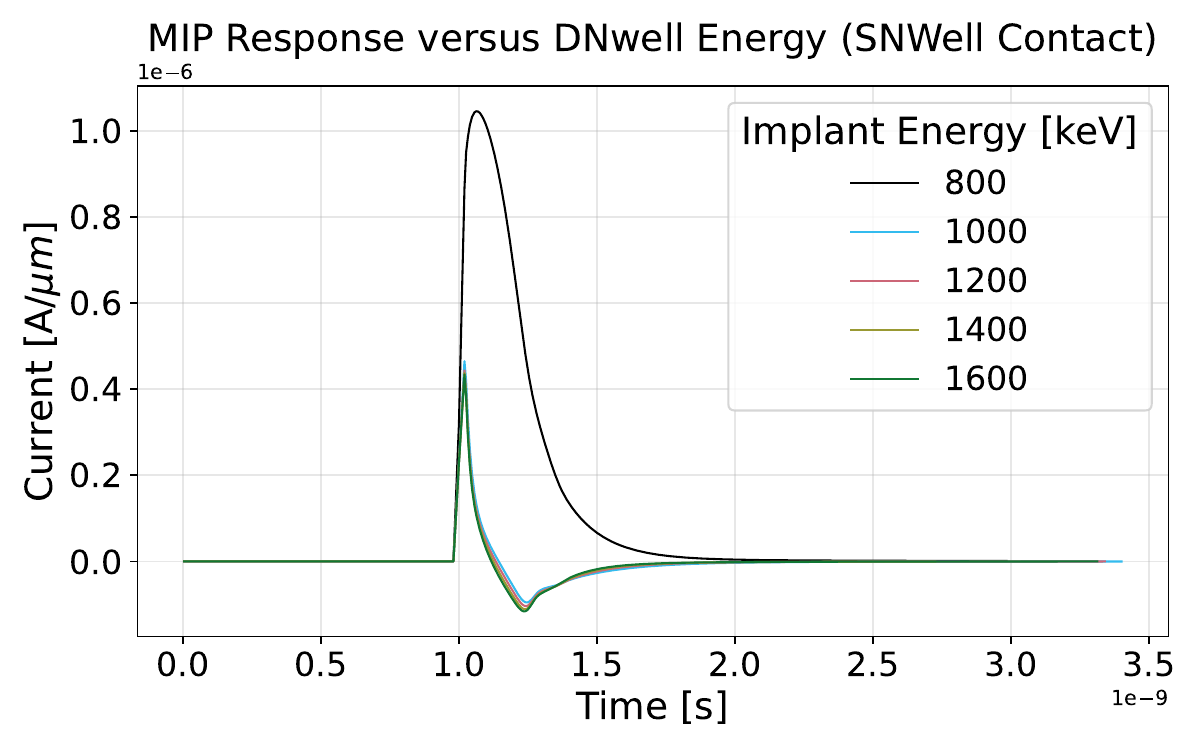}
    \includegraphics[width=0.32\textwidth]{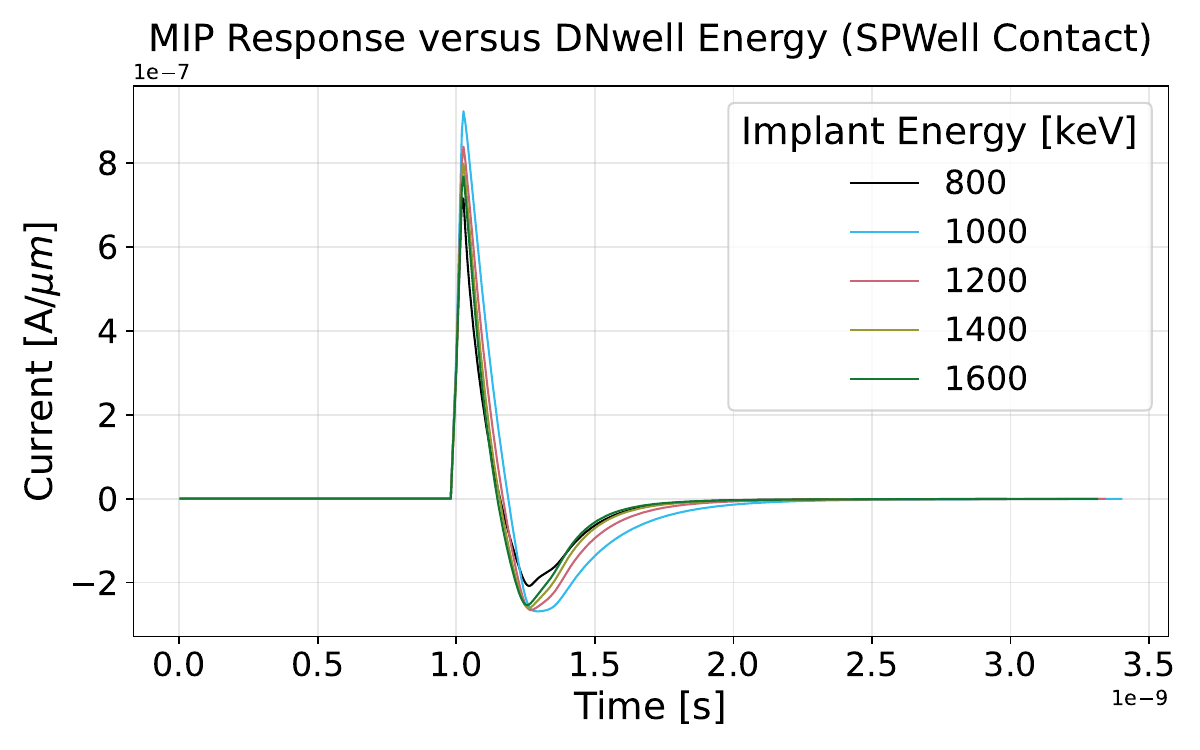}
    \caption{Response of the (left) DNwell, (center) SNwell, and (right) SPwell contacts to normally incident MIPs traversing the center of the pixel, shown as a function of the DNwell implantation energy.}
    \label{fig:mip-response-vs-energy}
\end{figure}

A complete presentation of the results from the optimization studies described above would be lengthy and largely redundant.
Therefore, only the results of the DNwell optimization are presented here, as similar trends were observed for the other top-side implants.
Figures~\ref{fig:mip-response-vs-dose}~and~\ref{fig:mip-response-vs-energy} provide representative examples that illustrate the observed trends.

The currents induced on the three implants, corresponding to the DNwell, SNwell, and SPwell, are shown as functions of the DNwell implantation dose in Figure~\ref{fig:mip-response-vs-dose}.
The signal amplitude at the DNwell contact initially increases with implantation dose, but begins to decrease for doses $\geq 3.5\times10^{13}~\unit{\centi\meter^{-2}}$.
In this regime, an increasing fraction of the signal is collected at the SNwell contact.
This behavior is understood to result from increased overlap between the DNwell and Piso layers, which reduces the effectiveness of the intended isolation.

The currents induced on the three implants, corresponding to the DNwell, SNwell, and SPwell, are shown as functions of the DNwell implantation energy in Figure~\ref{fig:mip-response-vs-energy}.
The signal amplitude at the DNwell contact generally increases with implantation energy.
However, for an implantation energy of 800~\unit{\kilo\eV}, almost no signal is induced at the DNwell contact, and most of the charge is collected at the SNwell contact.
At low implantation energies, the DNwell layer begins to overlap with the Piso layer, causing a majority of the charge to be collected at the SNwell contact.

\begin{figure}[hbtp]
    \centering
    \includegraphics[width=0.49\textwidth]{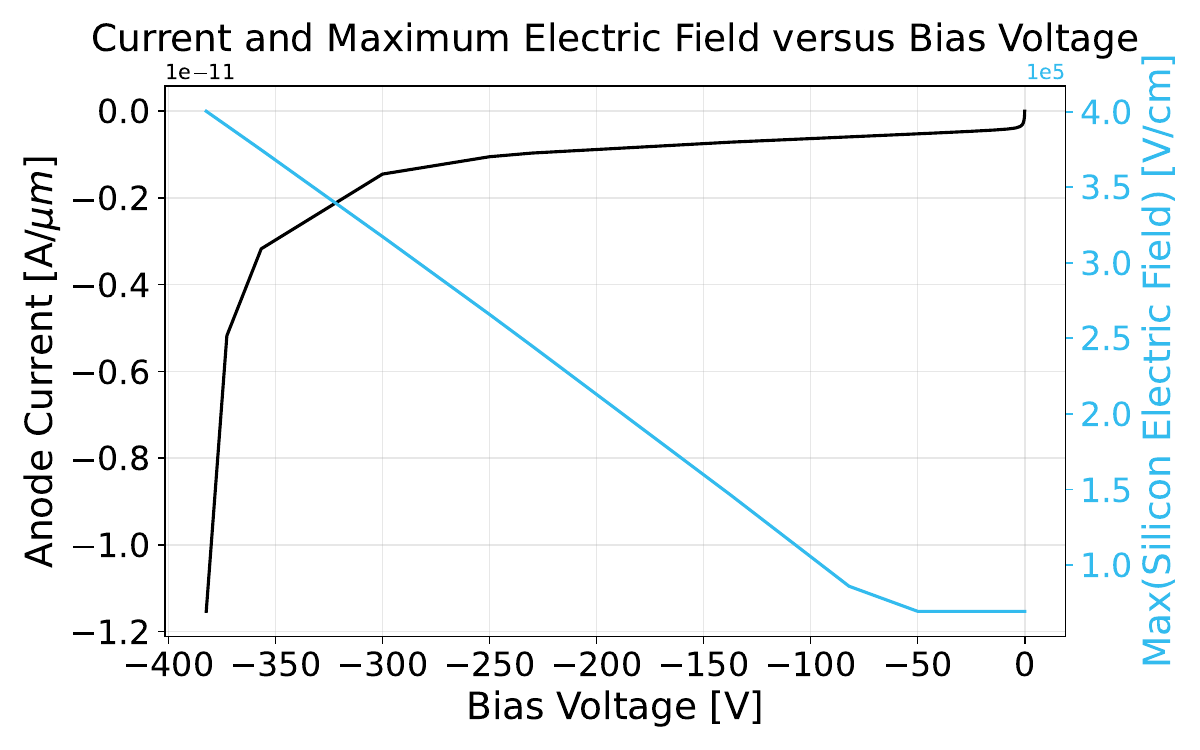}
    \includegraphics[width=0.49\textwidth]{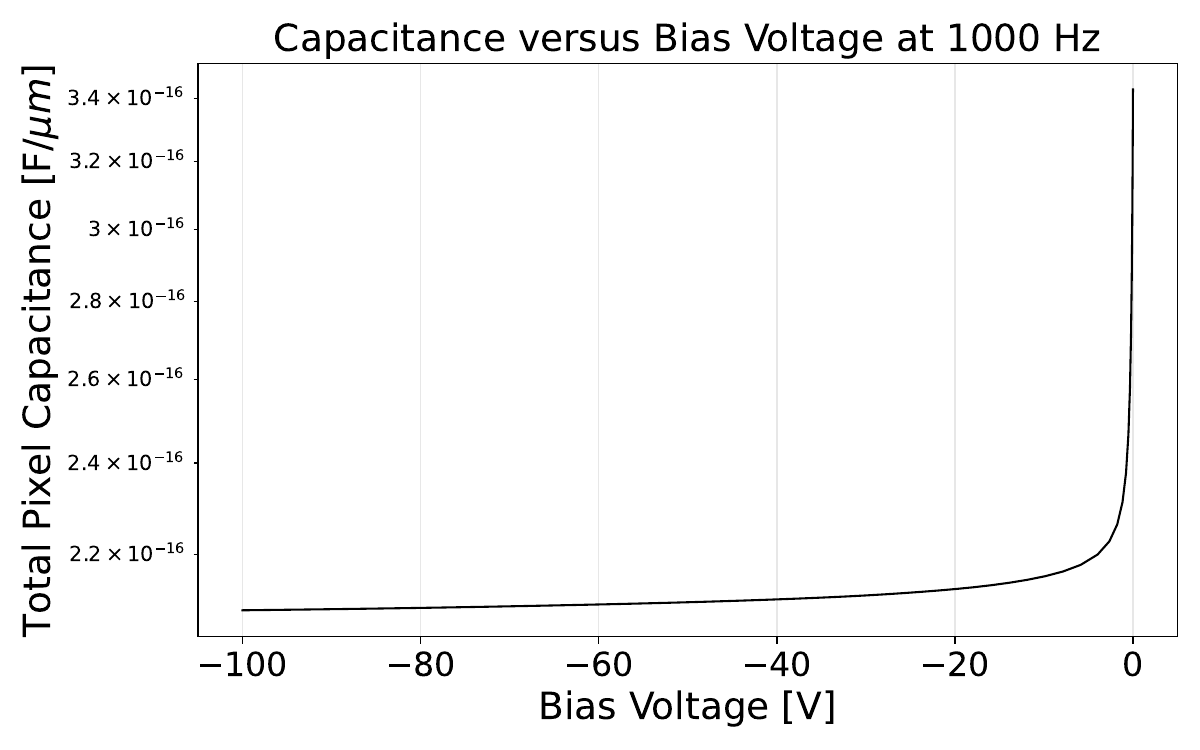}
    \caption{(left) Total current and maximum electric field within the semiconductor volume as functions of the bias voltage. (right) Total pixel capacitance as a function of the bias voltage, showing the saturation of the capacitance near the chosen operating voltage of $-100$~\unit{V}.}
    \label{fig:iv-cv-curves}
\end{figure}

The total current and maximum electric field magnitude as functions of the bias voltage for the final design are shown in Figure~\ref{fig:iv-cv-curves} (left).
The magnitude of the full depletion voltage is observed to be approximately 3~V, compared with a theoretical estimate of 0.5~V obtained from
\begin{equation}
V_{dep} = \frac{q|N_{eff}|d^2}{2\epsilon_{Si}},
\end{equation}
where $q$ is the elementary charge, $N_{eff}$ is the effective doping concentration of the silicon bulk, $d$ is the active thickness of the sensor, and $\epsilon_{Si}$ is the absolute permittivity of silicon.

The total pixel capacitance as a function of the bias voltage is shown in Figure~\ref{fig:iv-cv-curves} (right).
The magnitude of the breakdown voltage is estimated to be between 300 and 350~\unit{V}, assuming a breakdown electric field for bulk silicon of approximately $3\times10^{5}~\unit{V}/\unit{cm}$.
A complementary study, in which breakdown is identified as the bias voltage at which the ionization integral reaches unity, yields an estimated breakdown voltage magnitude of approximately 390~\unit{V}.

\begin{figure}[hbtp]
    \centering
    \includegraphics[width=0.5\linewidth]{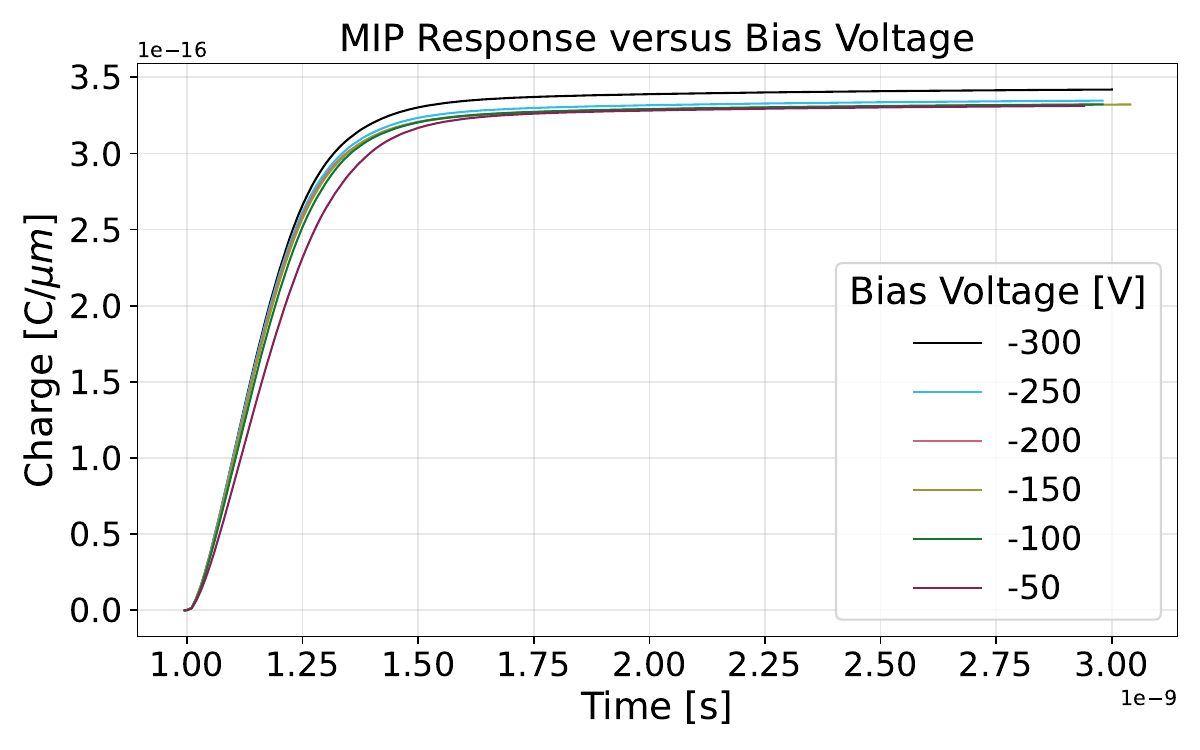}
    \caption{Integrated charge collected as a function of time for different bias voltages.
    The MIP traverses the center of the pixel at normal incidence, depositing 80 electron-hole pairs per~\unit{\micro\meter}.}
    \label{fig:charge-vs-bias-voltage}
\end{figure}

A bias voltage of $-100$~\unit{V} was chosen as a representative operating point.
Figure~\ref{fig:charge-vs-bias-voltage} shows the sensor response to MIPs traversing the center of the sensor at normal incidence as a function of the bias voltage, demonstrating that the response varies only modestly from that at $-100$~\unit{V} over the range studied.

\section{Simulated performance of HV-MAPS design}\label{sec:hvmaps-performance}
The charge-collection efficiency (CCE) was evaluated for normally incident charged particles at varying impact positions, as well as for particles incident at varying angles through the center of the pixel (\textit{i.e.}, $X = 0$~\unit{\micro\meter}).
The incident angle is defined with respect to the X-axis, such that $-90^\circ$ corresponds to a trajectory parallel to the Y-axis and $0^\circ$ corresponds to a trajectory parallel to the X-axis.
The CCE is defined either for the leading pixel or as the sum over all pixels.

The leading-pixel CCE ("Leading DNwell CCE") is defined as the ratio of the charge collected by the DNwell electrode with the largest signal to the total charge deposited by the MIP.
The summed CCE ("Sum(DNwell)") is defined as the ratio of the charge collected by all DNwell electrodes to the total charge deposited by the MIP, thereby accounting for charge sharing.

\begin{figure}[hbtp]
    \centering
    \includegraphics[width=0.49\textwidth]{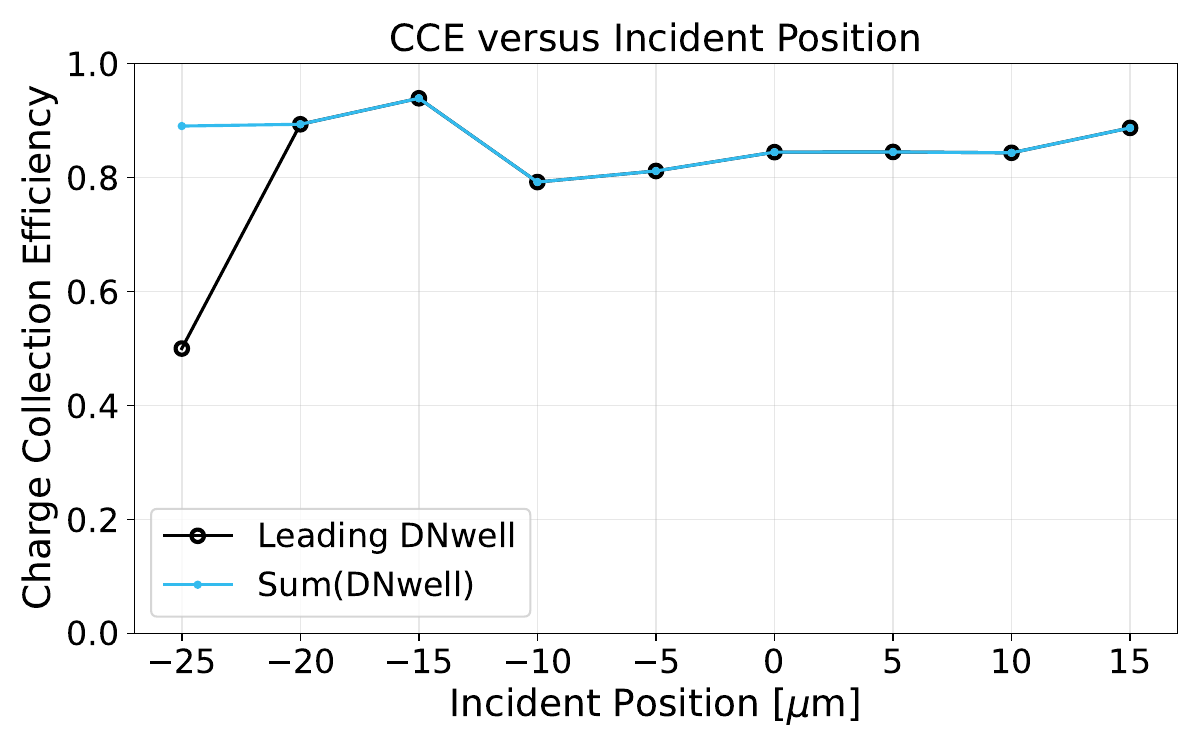}
    \includegraphics[width=0.49\textwidth]{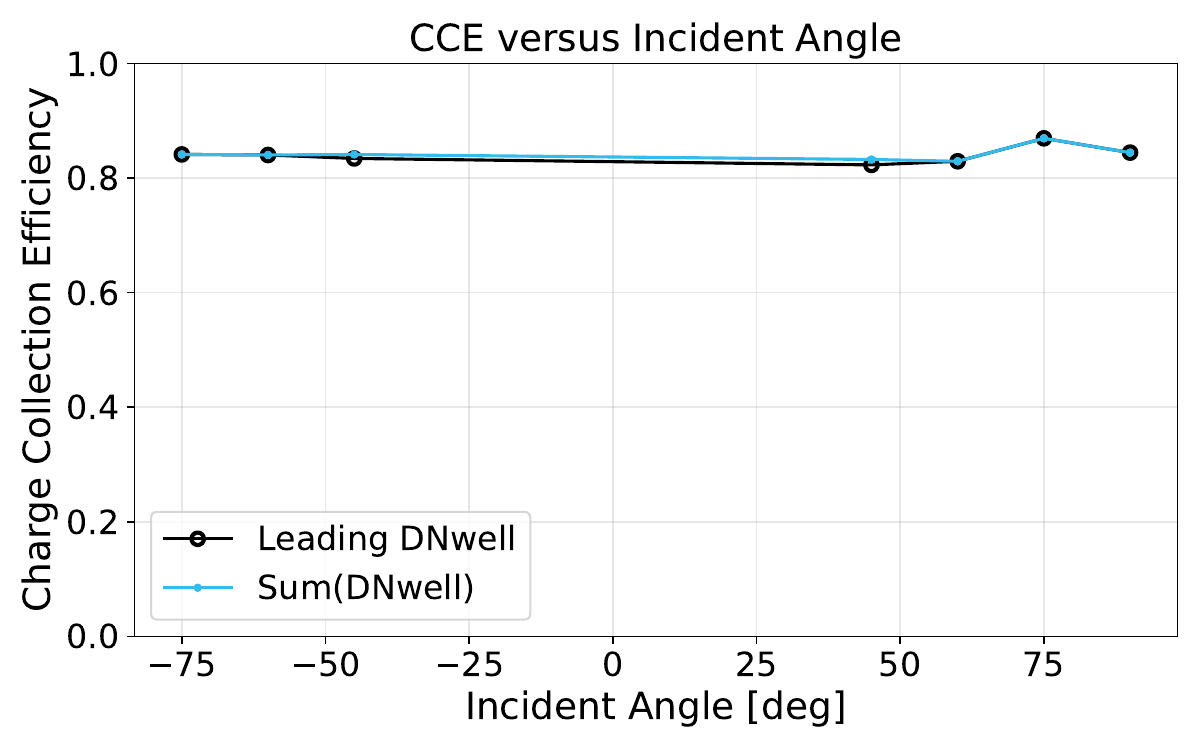}
    \caption{Charge-collection efficiency as a function of the MIP (left) incident position and (right) incident angle.
    The MIPs are normally incident when the incident position is varied and traverse the center of the pixel when the incident angle is varied.}
    \label{fig:cce-results}
\end{figure}

Figure~\ref{fig:cce-results} (left) shows the CCE as a function of the incident position across the central pixel.
Due to the relatively low electric field over a significant fraction of the DNwell volume, charge deposited in these regions is not fully collected.
This results in a CCE of approximately 80\% for tracks traversing the well regions, compared with approximately 90\% elsewhere.
The leading DNwell CCE decreases near an incident position of $-25$~\unit{\micro\meter} because the deposited charge is shared between neighboring pixels.

The dependence of the CCE on the incident angle was also studied, with the results shown in Figure~\ref{fig:cce-results} (right).
The CCE remains approximately constant with angle for the chosen incident position, although a small difference is observed between the values at $-75^\circ$ and $+75^\circ$.
This difference arises from the distinct material traversed by the charged particle: at $-75^\circ$, the particle passes through the DNwell, Piso, and SNwell regions, resulting in more complex charge collection than at $+75^\circ$, where it traverses the DNwell and SPwell regions.

\section{Conclusion and outlook}\label{sec:conclusion}

The design of an HV-MAPS pixel structure for the vertex detectors of \FCCee experiments was presented.
The sensor has a thickness of 25~\unit{\micro\meter}, a pixel pitch of 50~\unit{\micro\meter}, and a bulk resistivity of 13.3~\unit{\kilo\ohm\cdot\cm}.
A deep n-type well serves as the collection electrode, while a p-type isolation layer separates it from the shallow wells that will contain the in-pixel CMOS circuitry.
The implantation energy and dose of each top-side implant were optimized using two-dimensional TCAD simulations.
The DNwell scans illustrate the principal design trade-off: an excessively high implantation dose or an excessively low implantation energy causes the DNwell to overlap with the Piso layer.
This overlap degrades the isolation and diverts a significant fraction of the signal to the shallow n-type well contact.
Similar trends were observed for the other top-side implants.

The optimized design fully depletes at approximately $-3$~\unit{V} and has an estimated breakdown voltage magnitude of 300--390~\unit{V}, providing a wide operating-voltage range.
A bias voltage of $-100$~\unit{V} was chosen for the sensor-performance simulations.
At this bias, the simulated charge-collection efficiency is approximately 90\% for tracks outside the well regions and approximately 80\% for tracks traversing them, where the electric field is weaker.
The efficiency is nearly independent of the angle of incidence, apart from a small asymmetry caused by the different implants traversed at positive and negative angles.

Further steps are planned to advance the design toward fabrication of test structures.
These include replacing the simplified implant structures with detailed in-pixel CMOS circuitry, studying the effects of radiation-induced damage, and performing three-dimensional simulations to obtain more precise estimates of spatial resolution and charge sharing.
Once the design is finalized, test structures will be fabricated at a foundry and characterized through laboratory and test-beam measurements.

\acknowledgments

 This manuscript is based upon work supported by Laboratory Directed Research and Development (LDRD) funding from Fermi National Laboratory, provided by the Director, Office of Science, of the U.S. Department of Energy under Contract No. 89243024CSC000002. The authors also acknowledge support from the University of Illinois Chicago at Chicago and the U.S. National Science Foundation (NSF).


\providecommand{\href}[2]{#2}\begingroup\raggedright\endgroup

\end{document}